\documentclass{PoS}
\PoS{PoS(LAT2005)071}
\title{Search for the S=+1 pentaquarks in quenched lattice QCD }
\ShortTitle{Search for the S=+1 pentaquarks in quenched lattice QCD }

\author{\speaker{Toru T. Takahashi}\\
        Yukawa Institute for Theoretical Physics, Kyoto university, Kyoto 606-8502, Japan\\
        E-mail: \email{ttoru@yukawa.kyoto-u.ac.jp}}
\author{{Takashi Umeda}\\
        Yukawa Institute for Theoretical Physics, Kyoto university, Kyoto 606-8502, Japan\\
        E-mail: \email{tumeda@yukawa.kyoto-u.ac.jp}}
\author{{Tetsuya Onogi}\\
        Yukawa Institute for Theoretical Physics, Kyoto university, Kyoto 606-8502, Japan\\
        E-mail: \email{onogi@yukawa.kyoto-u.ac.jp}}
\author{{Teiji Kunihiro}\\
        Yukawa Institute for Theoretical Physics, Kyoto university, Kyoto 606-8502, Japan\\
        E-mail: \email{kunihiro@yukawa.kyoto-u.ac.jp}}

\abstract{
We study spin $\frac12$ hadronic states in
quenched lattice QCD to search for a possible $S=+1$ pentaquark
resonance. Our work is the first systematic lattice QCD study 
which properly carries out the following analyses:
{\it (1) the careful extraction of the first two low energy states 
with very high statistics and the variational method}
and
{\it (2) the study of volume dependences of eigenenergies and spectral 
weights to distinguish resonance states from scattering states}.

Simulations are carried out on 
$8^3\times 24$, $10^3\times 24$, $12^3\times 24$
and $16^3\times 24$ lattices at $\beta$=5.7
with the standard plaquette gauge action and the Wilson quark action.
Our result indicates the existence of a resonance state 
lying slightly above the NK threshold in $(I,J^P)=(0,\frac12^-)$ channel
in quenched QCD.
}
\FullConference{XXIIIrd International Symposium on Lattice Field Theory\\
		 25-30 July 2005\\
		 Trinity College, Dublin, Ireland}

\begin{document}

\section{Introduction}

After the first discovery~\cite{Netal03} of $\Theta^+(1540)$,
identifying the properties of the particle is one of the central problems
in hadron physics.
While the isospin of $\Theta^+$ is likely to be zero,
the spin and the parity and the origin
of its tiny width still remain open questions.
In spite of many theoretical and experimental
studies on $\Theta^+$~\cite{Hicks},
the nature of this exotic particle,
including the very existence of the particle, is {\it still} controversial.
In such a situation, 
the lattice QCD calculation is considered as one of the
most reliable methods, which is an {\it ab initio} method directly based on QCD.

Up to now, several lattice QCD studies have been reported,
which mainly look for pentaquarks in various different ways
~\cite{LattPenta,TTTT0405}.
However, the conclusion has not been unfortunately settled yet.
The difficulties in the spectroscopy calculation with 
lattice QCD generally arise from
systematic errors due to the discretization, the chiral extrapolation, the 
quenching effect, the finite volume effect and the contaminations 
from higher excited-states. 
The difficulty specific to the current problem 
is that the signal of $\Theta^+$ is embedded 
in the discrete spectrum of NK scattering states in finite volume. 
In order to verify the existence of a resonance state, 
one needs to isolate the first few low energy states including the lowest 
NK scattering state, identify a resonance state and study its 
volume dependence which can distinguish itself from other scattering 
states. 
Ideally, one should extract multistates from a high 
statistics unquenched calculation for several different physical 
volumes, where both the continuum and the chiral limits are taken.
However, there are no lattice QCD study which performs all these steps
so far, due to the enormous computational costs needed.
We therefore concentrate ourselves on the analyses using rather heavy 
quarks on coarse quenched lattices 
but with a good statistics and a proper separation of states.

In this report, we study $(I,J)=(0,\frac12)$ channel in quenched 
lattice QCD to search for possible resonance states.
We prepare thousands of gauge configurations in order to
single out the signals of a resonance state with good statistics.
We adopt two independent operators with $I=0$ and $J=\frac12$
and diagonalize the $2\times 2$ correlation matrices 
for all the combinations of lattice sizes 
and quark masses to extract the 2nd-lowest state slightly 
above the NK threshold in this channel.
After the careful separation of the states, we investigate the 
volume dependence of the energy as well as the spectral weight~\cite{Metal05}
of each state so that we can distinguish 
the resonance signal from the background signals of NK scattering states.
Our calculation
(For the details, please see our recent paper, Ref.~\cite{TTTT0405}.)
gives the first systematic study which performs
\begin{itemize}
\item
the analysis using correlation matrices with very high statistics,
which enables us 
to isolate the first few low energy states including the lowest 
NK scattering state
\item
the study of the volume dependences of eigenenergies and 
spectral weight factors with several different physical volumes,
which enables us to distinguish a resonance state
from scattering states
\end{itemize}
in a proper way.

\section{Setup}

We adopt the variational method using 
correlation matrices constructed from independent operators~\cite{TTTT0405}
in order to isolate the signal of a resonance state from the 
NK scattering states.
The simulations are carried out on four different sizes of lattices,
$8^3\times 24$, $10^3\times 24$, $12^3\times 24$ and $16^3\times 24$
with 2900, 2900, 1950 and 950 gauge configurations, respectively,
using the standard 
plaquette (Wilson) gauge action at $\beta=5.7$ and the Wilson quark action. 
The hopping parameters for the quarks are 
$(\kappa_{u,d},\kappa_{s})$=$(0.1600,0.1650)$, $(0.1625,0.1650)$,
$(0.1650,0.1650)$, $(0.1600,0.1600)$ and $(0.1650, 0.1600)$,
which correspond to the current quark masses 
$(m_{u,d},m_{s})\sim(240,100)$, $(170,100)$,
$(100,100)$, $(240,240)$ and $(100,240)$, respectively in the unit of MeV.
The lattice spacing $a$ determined from the Sommer scale
is about 0.17 fm, which implies the physical lattice sizes are
$1.4^3\times 4.0$ fm$^4$, $1.7^3\times 4.0$ fm$^4$, 
$2.0^3\times 4.0$ fm$^4$ and $2.7^3\times 4.0$ fm$^4$.
For the correlation matrices, we adopt two wall-type operators
$\overline{\Theta}^1_{\rm wall}(t)$ and $\overline{\Theta}^2_{\rm wall}(t)$
~\cite{TTTT0405}, whose spinor and color structures are the same as
those adopted in the first paper by Csikor {\it et al.}.
Periodic boundary conditions are taken in all directions for the gauge field,
whereas we impose periodic boundary conditions on the spatial directions
and the Dirichlet boundary condition on the temporal direction 
for the quark field
in order to avoid possible contaminations from those propagating
beyond the boundary at $t=0$ in (anti)periodic boundary 
conditions~\cite{TTTT0405}, which is peculiar to pentaquarks
and has not been dwelled.

\section{Lattice QCD Result in $(I,J^P)=(0,\frac12^-)$ channel}
\subsection{volume dependence of eigenenergies}

\begin{figure}[h]
\begin{center}
\includegraphics[scale=0.3]{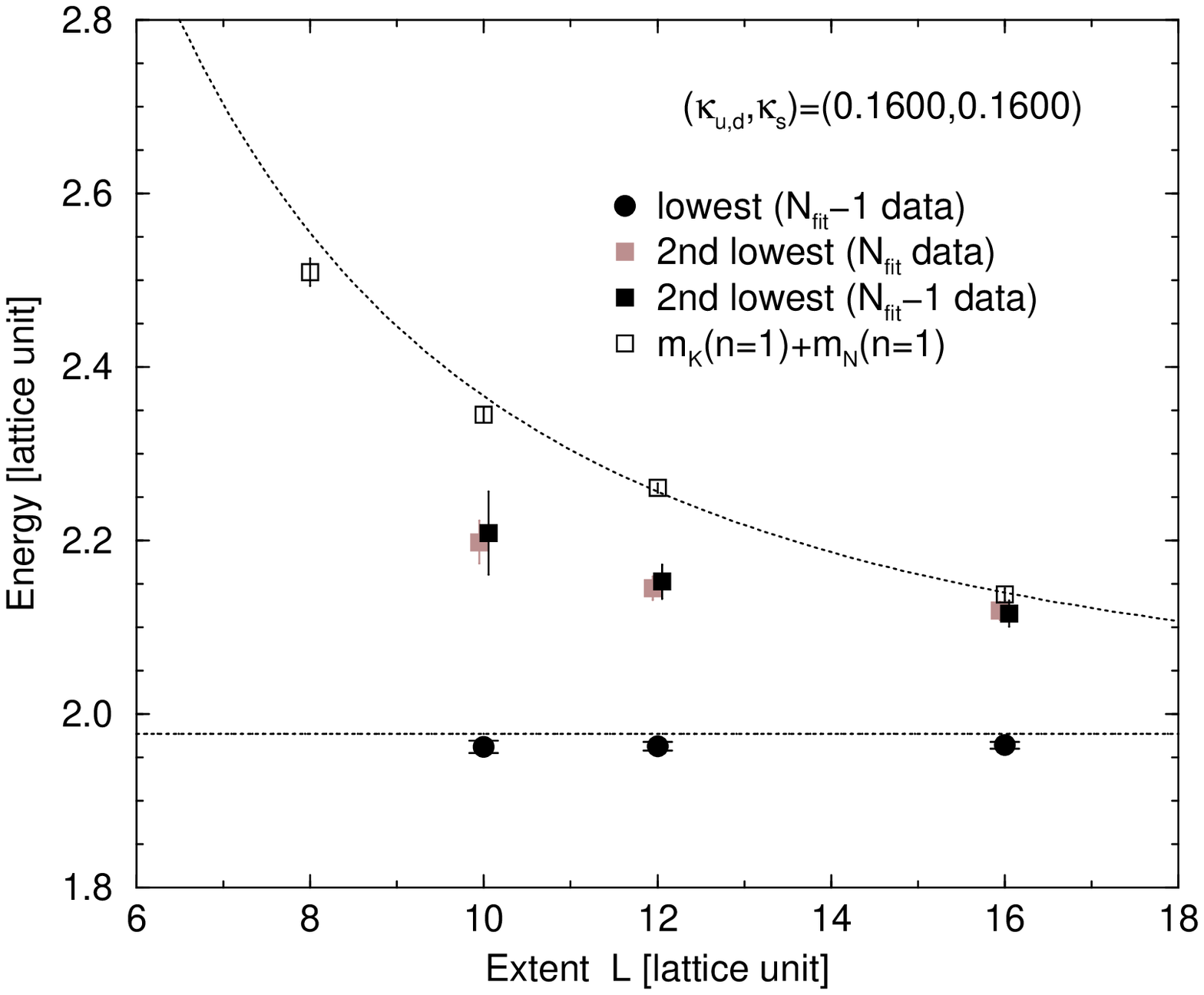}
\includegraphics[scale=0.3]{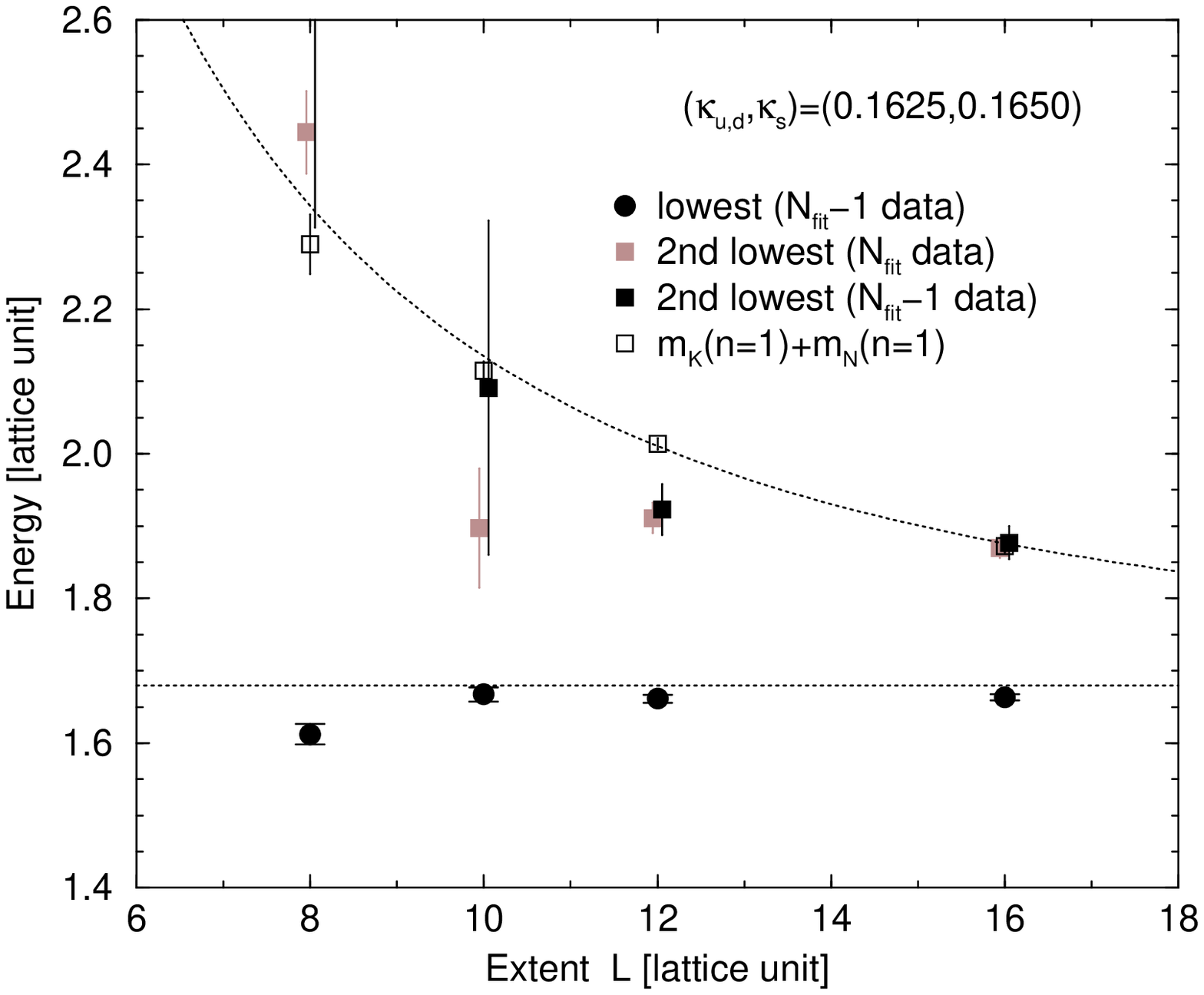}
\includegraphics[scale=0.3]{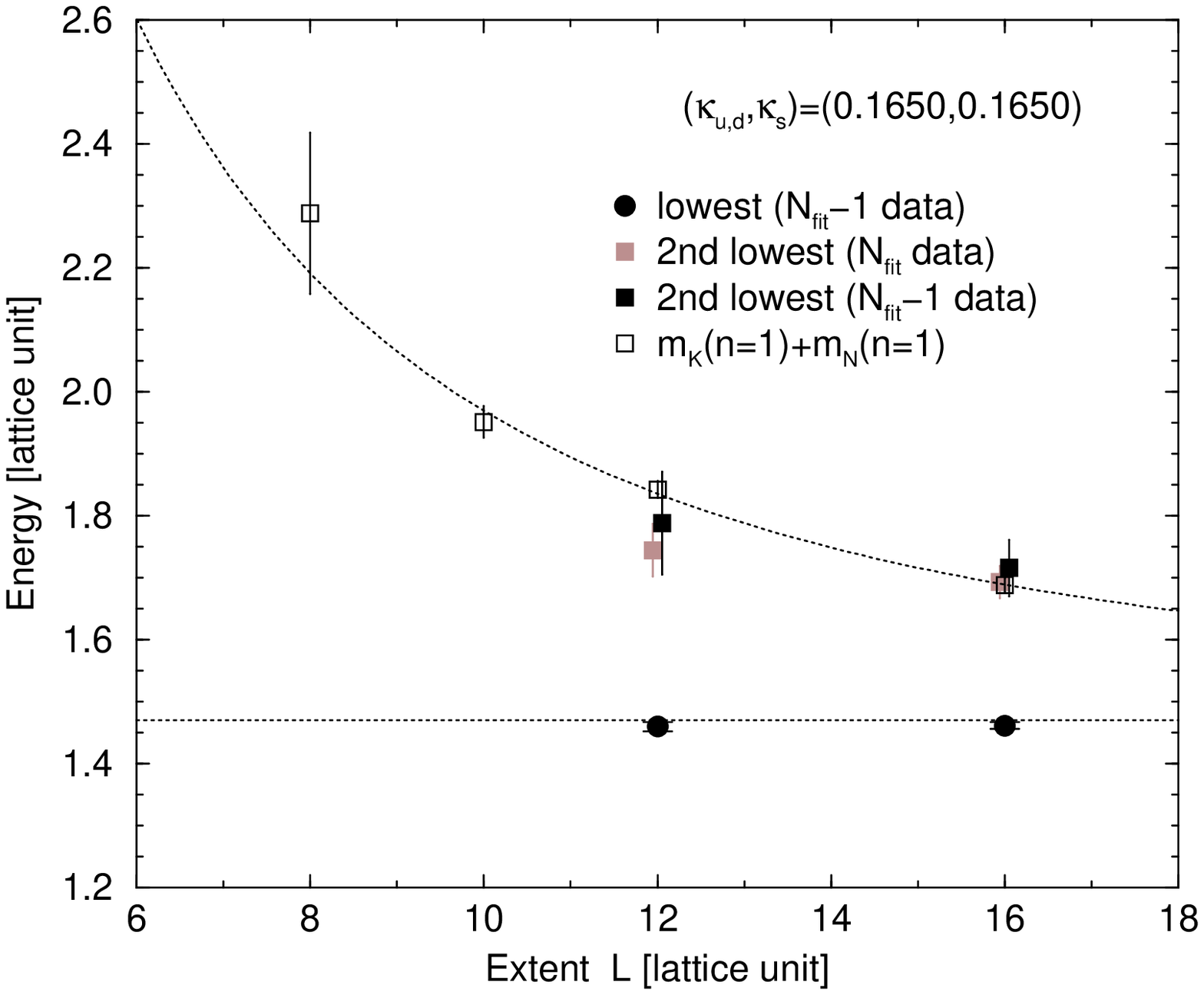}
\end{center}
\caption{\label{negativeGSES}
The black (gray) filled-squares denote
the lattice QCD data of the 2nd-lowest state in
$(I,J^P)=(0,\frac12^-)$ channel
extracted with $N_{\rm fit}-1$ ($N_{\rm fit}$) data 
plotted against the lattice extent $L$.
The filled circles represent 
the lattice QCD data $E_0^-$ of the lowest state in
$(I,J^P)=(0,\frac12^-)$ channel.
The open symbols are the sum 
$E_N^{\rm |\vec{\bf n}|=1} + E_K^{\rm |\vec{\bf n}|=1}$
of energies of nucleon $E_N^{\rm |\vec{\bf n}|=1}$
and Kaon $E_K^{\rm |\vec{\bf n}|=1}$ with the
smallest lattice momentum 
$|\vec {\bf p}|=\frac{2\pi}{L}|\vec{\bf n}|=2\pi/L$.
The upper line represents
$\sqrt{M_N^2+|{\bf p}|^2}+\sqrt{M_K^2+|{\bf p}|^2}$
with $|{\bf p}|=2\pi/L$ the smallest relative momentum
on the lattice.
The lower line represent the simple sum $M_N+M_K$
of the masses of nucleon $M_N$ and Kaon $M_K$.
We adopt the central values of $M_N$ and $M_K$ 
obtained on the largest lattice to draw the two lines.
}
\end{figure}

The filled circles in
Fig.~\ref{negativeGSES} show the lowest-state energies $E_0^-$
in $I=0$ and $J^P=\frac12^-$ channel on four different volumes.
Here the horizontal axis denotes the lattice extent $L$
in the lattice unit and the vertical axis represents the energy of the state.
The lower line denotes the simple sum $M_N+M_K$
of the nucleon mass $M_N$ and Kaon mass $M_K$
obtained with the largest lattice.
We here simply use $M_N+M_K\ (L=16)$ as a guideline.

As for the lowest state of this channel,
we find that the energy of the state takes almost constant value against
the volume variation and coincides with
the simple sum $M_N+M_K$. It is then concluded that 
the lowest state in $I=0$ and $J^P=\frac12^-$ channel
is the NK scattering state with the relative momentum $|{\bf p}|=0$.
The good agreement with the sum $M_N+M_K$ implies
the weakness of the interaction between N and K. 

The $(I,J^P)=(0,\frac12^-)$ state is one of the candidates
for $\Theta^+(1540)$.
Since $\Theta^+(1540)$ is located above the NK threshold,
it would appear as an excited state in this channel.
The volume dependence of the energy of each state can be used
to distinguish a possible resonance state from NK scattering states;
it is expected that the energies of resonance states 
have small volume dependence, whereas
the energies of NK scattering states are expected to scale
according to the lattice size $L$.

A possible candidate
for the volume dependence of the energies of NK scattering states
is the simple formula as
$E_{\rm NK}^{\vec{\bf n}}(L)\equiv
\sqrt{M_N^2+|\frac{2\pi}{L}\vec{\bf n}|^2}+
\sqrt{M_K^2+|\frac{2\pi}{L}\vec{\bf n}|^2}$ with
the relative momentum $\frac{2\pi}{L}\vec{\bf n}$ between N and K in
finite periodic lattices. There may be
some corrections to $E_{\rm NK}^{\vec{\bf n}}(L)$ in practice.
We therefore estimate three possible corrections~\cite{TTTT0405};
the existence of the NK interaction,
the application of the momenta on a finite discretized lattice and
the estimation of the implicit finite-size effects.
We find that these corrections lead to at most a few \% 
deviations from $E_{\rm NK}^{\vec{\bf n}}(L)$ and we
neglect these corrections for simplicity 
and use the simple form $E_{\rm NK}^{\vec{\bf n}}(L)$
in the following discussion.

We now compare the lattice data $E_1^-$ with
the expected behaviors $E_{\rm NK}^{\rm |\vec{\bf n}|=1}$
for the 2nd-lowest NK scattering states.
The filled squares in Fig.~\ref{negativeGSES}
denote $E_1^-$, the 2nd-lowest-state energies in this channel.
(The black and gray symbols are the lattice data obtained by the fits with 
two different fit ranges,
which help the readers to see the fit-range dependences.)
The upper line shows $E_{\rm NK}^{\rm |\vec{\bf n}|=1}$ estimated
with the next-smallest relative momentum between N and K,
and with the masses $M_K$ and $M_N$ extracted on the $L=16$ lattices.
Although the lattice QCD data $E_1^-$ and 
$E_{\rm NK}^{\rm|\vec{\bf n}|=1}$
almost coincide with each other on the $L=16$ lattices, 
which one may take as the characteristics of
the 2nd-lowest scattering state,
the data $E_1^-$ do not follow $E_{\rm NK}^{\rm|\vec{\bf n}|=1}$
in the smaller lattices.
Especially when the quarks are heavy,
composite particles would be compact objects and show smaller 
finite volume effects besides those arising from the lattice momenta 
$\vec{\bf p}=\frac{2\pi}{L}\vec{\bf n}$.
The statistical errors are also well controlled for the heavy quarks.
Thus, the significant deviations in $1.5\lesssim L\lesssim 3$ fm 
with the combination of the heavy quarks, 
such as $(\kappa_{u,d},\kappa_s)=(0.1600,0.1600)$,
are reliable and the obtained 2nd-lowest states
are difficult to be explained as the NK scattering states.
One can understand this behavior with the view
that the 2nd-lowest state 
is a resonance state rather than a scattering state.
In fact, while the data with the lighter quarks have rather strong volume
dependences which can be considered to arise because of the finite size of a
resonance state, the lattice data exhibit almost no volume dependence
with the combination of the heavy quarks
especially in $1.5\lesssim L\lesssim 3$ fm,
which can be regarded as the characteristic of resonance states.

\subsection{volume dependence of weightfactors}

For further confirmation, we investigate 
the volume dependence of the spectral weight~\cite{Metal05}.
In the case when a correlation function 
is constructed from a point-source and a zero-momentum point-sink,
the weight factor $W_i$ takes an almost constant value
if $|i\rangle$ is the resonance state where the wave function is localized.
If the state $|i\rangle$ is a two-particle state, the 
situation is more complicated. Nevertheless, when 
the interaction between the two particles is weak, 
which is the case in the present situation,
the weight factor is proportional to $\frac{1}{V}$.
When a source is a wall-type operator 
$\overline{\Theta}_{\rm wall}(t_{\rm src})$ as taken in this work, 
a definite volume dependence of $W_i$ is not known.
Therefore, we re-examine the lowest state and the 2nd-lowest state
in $(I,J^P)=(0,\frac12^-)$ channel 
with the hopping parameter $(\kappa_{u,d},\kappa_s)=(0.1600,0.1600)$
and the locally-smeared 
source $\overline{\Theta^{2}}_{\rm smear}(t_{\rm src})$~\cite{TTTT0405},
which is introduced to partially enhance the ground-state overlap.
We additionally employ $14^3\times 24$ lattice for this analysis.
\begin{figure}[h]
\begin{center}
\includegraphics[scale=0.25]{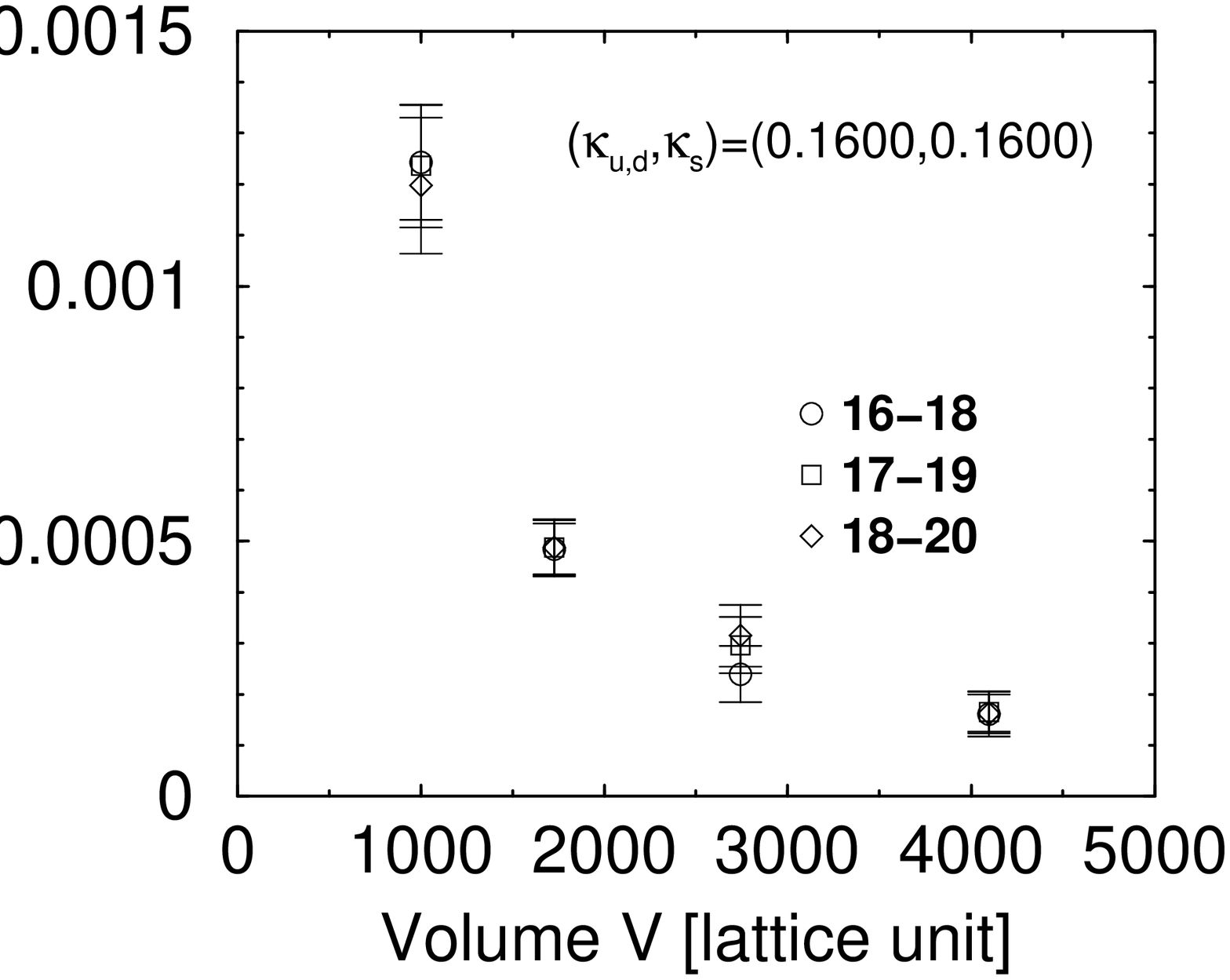}
\hspace{1cm}
\includegraphics[scale=0.25]{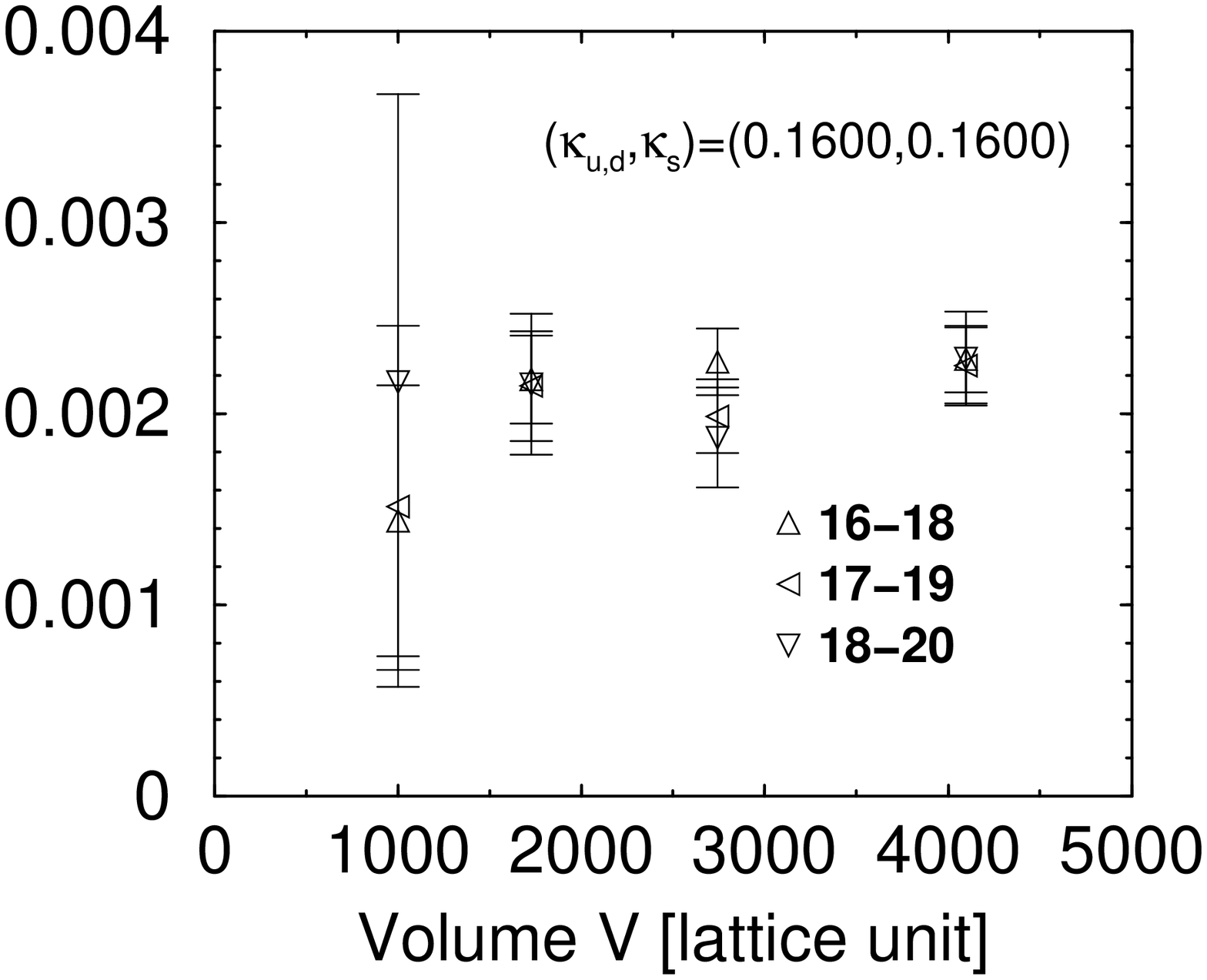}
\end{center}
\caption{\label{weightfactor}
The left figure shows $W_0$ for the lowest state in $(I,J^P)=(0,\frac12^-)$
channel and the right figure shows $W_1$ for the 2nd-lowest state.
In the case when the weight factor $W_i$ for the $i$-th state $|i\rangle$
in a point-point correlator
shows no volume dependence, $|i\rangle$ is likely to be a resonance state.
On the contrary, when the $i$-th state $|i\rangle$ is a two-particle
state, $W_i$ scales according to $1/V$.}
\end{figure}
We extract the weight factors $W_0$ and $W_1$ using the two-exponential fit
as\\ $\sum_{\vec{\bf x}}\langle\Theta^2(\vec{{\bf x}},T+t_{\rm src})
\overline{\Theta^{2}}_{\rm smear}(\vec{{\bf 0}},t_{\rm src})\rangle
=W_0e^{-E_0^2T}+W_1e^{-E_1^-T}$,
with the exponents $E_0^-$ and $E_1^-$ fixed.
We try the various fit ranges as ($T_{\rm min}$,$T_{\rm max}$)=
(16,18),(17,19),(18,20) to see the fit-range dependences.
Fig.~\ref{weightfactor} includes all the results and
one can find that the global behaviors are almost the same among these three.

The figures in Fig.~\ref{weightfactor} show
the weight factors $W_0$ and $W_1$
of the lowest and next-lowest state in $(I,J^P)=(0,\frac12^-)$
channel, respectively.
We find that the 
volume dependence of $W_0$ on $V$ is consistent with $\frac{1}{V}$,
which is expected in the case of two-particle states.
On the other hand,
$W_1$ shows almost no volume dependence against $V$,
which is the characteristic of the state
in which the relative wave function is localized.
This result can be considered as another evidence
of a resonance state lying slightly above the NK threshold.

\section{Lattice QCD Result in $(I,J^P)=(0,\frac12^+)$ channel}

%\begin{figure}[htb]
%\begin{center}
%\includegraphics[scale=0.25]{031031031_02_SS.eps}
%\includegraphics[scale=0.25]{033033031_02_SS.eps}
%\includegraphics[scale=0.25]{032032031_02_SS.eps}
%\includegraphics[scale=0.25]{031031032_02_SS.eps}
%\includegraphics[scale=0.25]{032032032_02_SS.eps}
%\end{center}
%\caption{\label{positiveSS}
%The lattice QCD data in the $(I,J^P)=(0,\frac12^+)$ channel
%are plotted against the lattice extent $L$.
%The solid line denotes the simple sum $M_{N^*}+M_K$ of
%the masses of the lowest-state negative-parity nucleon $M_{N^*}$
%and Kaon $M_K$ obtained with the largest lattice.
%}
%\end{figure}
In the same way as $(I,J^P)=(0,\frac12^-)$ channel,
we have attempted to diagonalize the correlation matrix
in $(I,J^P)=(0,\frac12^+)$ channel
using the wall-sources $\overline{\Theta}_{\rm wall}(t)$
and the zero-momentum point-sinks $\sum_{\vec{\bf x}}\Theta(\vec{\bf x},t)$.
In this channel, the diagonalization is rather unstable and we
find only one state except for tiny contributions of possible other states~\cite{TTTT0405}.
The data have almost no volume dependence
and they coincide with the solid line which represents the simple sum 
$M_{N^*}+M_K$ of $M_{N^*}$ and $M_K$,
with $M_{N^*}$ the mass of the ground state
of the {\it negative-parity} nucleon.
The state we observe is therefore concluded to be the $N^*$-$K$ 
scattering state with the relative momentum $|{\bf p}|=0$.
%It may sound strange
%because the p-wave state of N and K with the relative
%momentum $|{\bf p}|=2\pi/L$ should be lighter than
%the $N^*$-$K$ scattering state with the relative momentum $|{\bf p}|=0$
%for the present parameter region;
%this lighter state is missing in our analysis.
%This failure would be due to the wall-type operator $\Theta_{\rm wall}(t)$.
%The relation between operators and overlap coefficients
%is an interesting problem and is to be explored in detail
%for further studies.
%Anyway, the strong dependence on the choice of operators
%suggests that it is needed to try 
%various types of operators before giving the final conclusion.

\section{Summary}

We have studied the $(S,I,J)=(+1,0,\frac12)$ states
on $8^3\times 24$, $10^3\times 24$, $12^3\times 24$
and $16^3\times 24$ lattices at $\beta$=5.7 using quenched lattice QCD.
We have performed a systematic study by 
{\it (1) the variational analysis using correlation matrices
with very high statistics
and (2) the study of the volume dependences of eigenenergies and 
spectral weight factors with several different physical volumes}
in a proper way for the first time.
From the correlation matrix of the operators,
we have successfully obtained the energies of the lowest state and 
the 2nd-lowest state in the $(I,J^P)=(0,\frac12^-)$ channel.
The volume dependence of the energies and spectral weight factors
show that the 2nd-lowest state in this channel is likely to be
a resonance state located slightly above the NK threshold
and that the lowest state is the NK scattering state 
with the relative momentum $|{\bf p}|=0$.
As for the $(I,J^P)=(0,\frac12^+)$ channel,
we have observed only one state in the present analysis,
which is likely to be a $N^*K$ scattering state
of the ground state of the negative-parity nucleon $N^*$ and Kaon
with the relative momentum $|{\bf p}|=0$.

The chiral-extrapolated value of
the eigenenergy $E_1^-$ of the 2nd-lowest state in the negative parity channel
is $1.500(52)$ in the lattice unit,
which is about 120\% of the NK threshold $1.227(6)$.
This state is rather heavier than 
the mass of $\Theta^+(1540)$ in the real world.
One possibility is the systematic 
errors from the discretization, the chiral extrapolation, or quenching.
Another possibility is that the observed 2nd-lowest state might be a
signal of a resonance state lying higher than $\Theta^+$
in the quenched QCD.
It is difficult to give a clear explanation at this point and 
more extensive studies on finer lattices with lighter quark
masses in unquenched QCD will be required. However, we can at least 
conclude that our quenched lattice calculations suggest the 
existence of a resonance-like state slightly above the NK threshold 
for the parameter region we have investigated.

\section*{acknowledgments}

T.~T.~T. was supported by the Japan Society for the
Promotion of Science (JSPS) for Young Scientists. 
T.~O. and T.~K. are supported by Grant-in-Aid for Scientific research 
from the Ministry of Education, Culture, Sports, Science and
Technology of Japan (Nos. 13135213,16028210, 16540243)
and (Nos. 17540250), respectively. 
This work is also partially supported by the 21st Century for Center 
of Excellence program.  
The lattice QCD Monte Carlo calculations have been performed
on NEC-SX5 at Osaka University and on HITACHI-SR8000 at KEK.

\end{document}